\documentclass[12pt,preprint]{aastex6} 
\usepackage{subfigure}
\usepackage{graphicx}
\usepackage{amsmath}

\slugcomment{Draft, revised 1, \today}

\shorttitle{Can gyrochronological ages be affected by the variation of the magnetic braking index?}
\shortauthors{D. B. de Freitas}

\begin{document}

\title{Can gyrochronological ages be affected by the variation of the magnetic braking index?}

\author{D. B. de Freitas\altaffilmark{1}}
\altaffiltext{1}{Departamento de F\'{\i}sica, Universidade Federal do Cear\'a, Caixa Postal 6030, Campus do Pici, 60455-900 Fortaleza, Cear\'a, Brazil}

\begin{abstract}
The present work attempts to analytically explain the effects on the moment of inertia, $I$, caused by the evolution of stellar velocity in main-sequence stars. We have found that the effects linked to stellar oblateness can be estimated from the variation of the magnetic braking index denoted by $\Delta q$. We have also found that the effect of $\dot{I}$ is here a determining factor for understanding the delicate mechanisms that control the spin-down of stars in this evolutionary phase. We note that our models predict that the behaviour of the variations of the braking index is distinct when the star density decreases. In the present study, we used three sample stellar targets from the archives of the different satellites, such as CoRoT and \textit{Kepler}, to estimate the possible correlations among the parameters extracted from our model and the stellar age. As a result, we find that $\Delta q$ is strongly correlated to the age of stars. In conclusion, we suggest that the variations of the magnetic braking index due to oblateness can be an interesting way to estimate stellar ages.
\end{abstract}


\section{Introduction}
Understanding the longevity of systems on a human or cosmic scale is not only of interest to scientists but of interest to any ordinary people. Generally speaking, the mass of a star is a fundamental attribute for understanding its longevity in the Hertzsprung--Russell diagram. However, the mass of a star can only give overview of its lifetime.  Massive stars have shorter lives than the low-mass stars because their rate of nuclear reactions is very much greater and, therefore, hydrogen is depleted more quickly \citep{kipp}. 

Apart from its mass, the age of a star permits the study of the time evolution of astronomical phenomena, such as rotation, magnetic activity, and chemical abundances \citep{sku, reiners2012,max}. On the past several decades, a great deal of effort has been done on the possibility of using stars as clocks and, therefore, estimate their ages from rotational properties, one them, the rotation period \citep{barnes2007}. Still, there are different methods for estimating stellar ages based on the calibrated rotation period--age relationship \citep{barnes2010,max}. Nevertheless, these approaches have limitations. In particular, these methods do not account for changes in the moments of inertia, which are related to stellar oblateness. A particular motivation is to improve low-mass star age estimates that stems from moment of inertia evolution. The present paper is focused on this issue.

Since Isaac Newton's \textit{Principia}, it has been emphasized that even for a small rotation, a body must become slightly oblate in the direction of its axis of rotation. Furthermore, as mentioned by \cite{chandra1969}, Newton highlighted the equilibrium of the body demanding a simple proportionality between the effect of rotation, as measured by the oblateness, and its cause, as measured by the ratio between the centrifugal acceleration at the equator and the mean gravitational acceleration on the surface.
The small effects of oblateness appear in most astrophysics textbooks and papers as neglectable \citep{kipp}. In contrast, this effect carries very important information concerning the mechanism that governs stellar rotation and stellar age. 

\subsection{Proxies for age estimation} 
It is widely accepted that magnetic braking is a fundamental concept for understanding angular momentum losses due to magnetic stellar winds for several classes of stars, such as main-sequence field and cluster stars. This mechanism was initially suggested by \citet{Schatzman}, who pointed out that slow rotators have convective envelopes. As mentioned by \cite{Kraft}, the behaviour of the mean rotational velocity of low-mass-main-sequence stars below 1.5$M_{\odot}$ (spectral type F0) is preferentially due to magnetic wind. A few years later, \citet{sku}'s pioneering work argued that stellar rotation, activity, and lithium abundances for solar-like stars obeys= a simple relationship given by $\mathrm dJ/\mathrm dt \propto \Omega^{3}$, where $t$ is time, $J$ is the angular momentum and $\Omega$ denotes the angular velocity. These authors established early on that stellar rotation and age should be related in cool main-sequence stars. The next step was to develop a more complete theory capable of explaining how losses of angular momentum occurs.

Inspired by \cite{mestel1968,mestel1984,mestel1987}'works, \citet{kawaler1988} elaborated on a theoretical model for describing the behaviour of the loss of angular momentum for main-sequence stars with masses less than 1.5$M_{\odot}$ due to wind ejected by stars. This wind gets caught by the magnetic field that spins outward until it is ejected, affecting the angular momentum and causing slowdown. In this way, the magnetic field acts like a brake.

As the magnetic field strength depends on stellar mass, \citet{chaboyer1995} modified the Kawaler's parametrization and introduced a saturation level into the angular momentum loss law. On the other hand, \citet{kris1997} proposed the inclusion of a Rossby scaling at the saturation velocity for stars more massive than 0.5$M_{\odot}$. More recently, the Roosby number (defined as the ratio between the rotational period and the convective overturn timescale) is defined to characterize the deviation from the Skumanich law.

In this context, the rotation--age relationship offers a promising method for measuring the ages of field stars, providing the rotation as an attractive alternative chronometer \citep{epstein2012}. Of the methods and recalibrations to estimate the ages of stars, the present work will focus on three: gyrochronology, isochrones and asteroseismology.

\cite{barnes2003} proposed a simple formulation that uses the colour and period values to derive the stellar ages for solar- and late-type stars. According to authors, the age and colour dependences of sequences of stars allow us to identify their underlying mechanism, which appears to be primarily magnetic. This determination of stellar ages from their rotational periods and colours, he named ``Stellar Gyrochronology''. The theoretical background is centred on a functional formulae based on the Skumanich-type age dependence fitted by the following expression\begin{equation}
\label{int1}
P=g(t)f(B-V)=\sqrt{t}f(B-V),
\end{equation} 
where $P$, $t$, and $B-V$ are the rotational periods (days), ages (Myr), and colours, respectively. However, only a sequence of stars, also called sequence $I$, respects this condition. This sequence consists of stars that form a diagonal band of increasing periods with increasing $B-V$ colours in a colour period diagram. There is also another sequence of stars denoted by the letter $C$ that represents the fast rotators. For these stars, the magnetic field is expected to be saturated. Consequently, there is no clear dependence between the period and colour. In this case, the expression for the angular momentum loss rate is given by $\mathrm dJ/\mathrm dt \propto \Omega$, and therefore, the period and age are related by a simple exponential law \citep{chaboyer1995,barnes2003,barnes2007,pacepas}.

Recently, \citet{defreitas2013} revisited the modified Kawaler parametrization proposed by \citet{chaboyer1995} in the light of the nonextensive statistical mechanics \citep{tsallis1988}. \citet{defreitas2013} analyzed the rotational evolution of the unsaturated F- and G- field stars that are limited in age and mass within the solar neighbourhood using a catalogue of $\sim$16000 stars in the main sequence \citep{holmberg2007}. They use the entropic index $q$ extracted from the Tsallis formalism as a parameter that describes the level of magnetic braking. They also linked this parameter with the exponent of dynamo theory ($a$) and with magnetic field topology ($N$) through the relationship $q=1+4aN/3$. As a result, they showed that the saturated regime can be recovered in the nonextensive context, assuming the limit $q\rightarrow1$. This limit is particularly important because it represents the thermodynamic equilibrium valid in the Boltzmannian regime. Indeed, the torque in the nonextensive version is given as $\mathrm dJ/\mathrm dt \propto \Omega^{q}$, revealing that the rotational velocities of F- and G-type main-sequence stars decrease with age according to $t^{1/(1-q)}$. The values of $q$ obtained by \citet{defreitas2013} suggest that it has a strong dependence on stellar mass.

In a recent paper, \cite{van} discuss the validity of the gyrochronological model for stars older than the Sun. They show that for ages greater than that of our Sun, a fundamental change occurs in the magnetic nature of the stars, revealing a critical transition that separates the strength of the stellar winds into two regimes. In short, the magnetic braking weakens as the star's age advances. With this knowledge, the authors point out that the use of gyrochronology for older stars in the main sequence becomes limited when weakened magnetic braking begins. To this end, they show that this behaviour occurs for a selected sample of 21 Kepler stars. In particular, this result reveals another important point for the present study: we cannot consider the same magnetic braking law for the lifetimes of the stars in the main sequence or even that stars must have the same magnetic braking mechanism as the Sun. Based on this assumption, the exponent $q$, which describes the evolution of stellar rotation, is not constant over time and acts according to the action of the stellar dynamic. In this sense, an alternative method considers a model that meets this prerogative and recalculate the ages in light of the variations of the magnetic braking index, denoted here by $\Delta q$. 

The main goal of the research is to show that rotation does not account for itself when explaining the star clock. It is necessary to verify a fine structure that can be responsible via the variation of the magnetic braking index marked by changes of the moment of inertia. Undoubtedly, the key feature here is the variation of the braking index, which cannot be held constant but instead evolves and diminishes on timescales shorter that the stellar life span. In this work, we present a new approach for understanding the behaviour of the magnetic braking index as a function of the change of the moment of inertia. 

Our paper is organized as follow. In Section 2, we give the main implications of the effects of changes of the moments of inertia for the magnetic braking index and briefly describe different physical mechanisms that could affect stellar rotation. In Section 3, after the investigation of the relations describing the variations of magnetic braking indexes, we show how the stellar age can be adjusted using the combined braking index $q$ and its variation $\Delta q$ as a function of rotation. In Section 4, we investigate a possible correlation between the variation of the magnetic braking index and asteroseismic ages. In the next section, we estimate the $\Delta q$-ages based on the power law proposed by \cite{defreitas2013} and the variations of magnetic braking. In Section 6, we test a set of stellar targets within the framework of our model and investigate the variations of the braking indexes as functions of the star's age and recalculate new ages based on the oblateness of the stars. Finally, in Section 5, conclusions are discussed.

\section{The effects of $\dot{I}$ on the variation of the magnetic braking index}

By using a nonextensive framework \citep{tsallis1998,tsallis1999,tsallis2004,defreitas2012,Silvaetal13}, \cite{defreitas2013} proposed the following equation to describe the behaviour of the rotational velocity of the main-sequence stars:
\begin{equation}
\begin{array}{cc}
\dot{\Omega}\propto \Omega^{^{q}}; & (q\geqslant 1).
\end{array}
\label{4}
\end{equation}

In the literature, the index $q$ is usually assumed to be a constant that measures the efficiency of magnetic braking during the lives of stars. This braking index can be define using the second derivate of $\Omega$ as a function of time. Considering that $q$ does not depend of time, it can be written as
\begin{equation}
q=\frac{\ddot{\Omega}\Omega}{\dot{\Omega}^{2}}.  \label{4.1}
\end{equation}

In the scenario which assumes spin-down due to stellar magnetic winds, the $q$-index is mostly 3, denoting by the Skumanich index. For main-sequence stars, eq. (\ref{4}) is more appropriated when the moment of inertia is a constant. 

However, the moment of inertia of a star can be affected by either rotation or stellar radius changes during its evolution. (In present study, the rate of mass loss is neglected). If the moment of inertia is affected by the rotation, we can approximate the star by an incompressible fluid and, therefore, its volume is constant. On the other hand, if the radius changes over time are taken into account, the volume of the star is not constant. 

In both cases, we can write the angular mometum loss rate as
\begin{equation}
\label{cap51}
\dot{J}=I\dot{\Omega} +\dot{I}\Omega.
\end{equation}

In the next section, we will investigate the effects of $\dot{I}$ when considering the star's volume as a constant, i.e., only the shape of the star changes. This case is particularly important for main-sequence stars.

\begin{figure*}
	\begin{center}
		\includegraphics[width=0.99\textwidth]{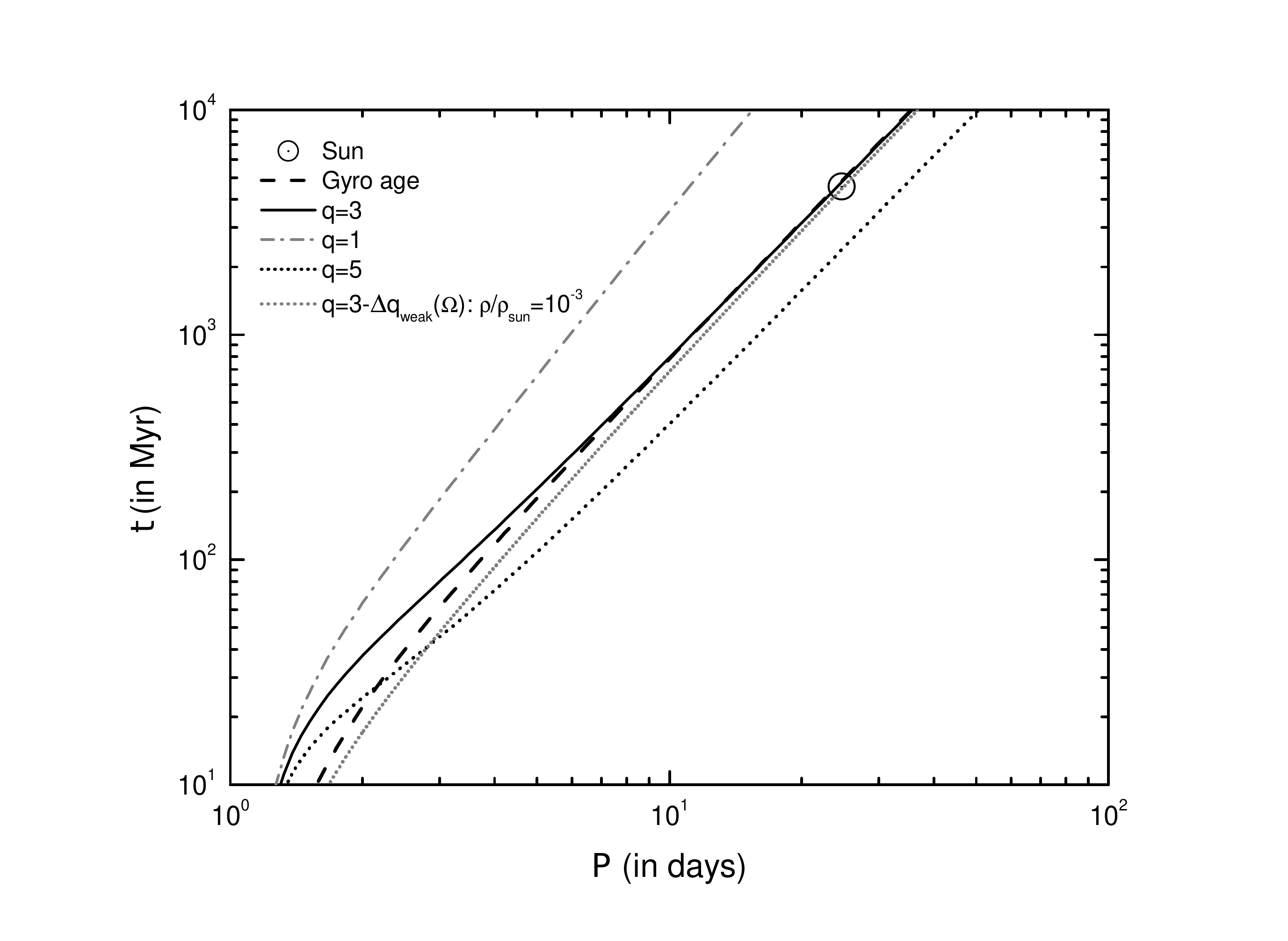}
	\end{center}
	\caption{Stellar age $t$ (in Myr) as a function of period $P$ (in days) for differing initial conditions of braking index $q$ (weak version) based on the density of stars. In particular, $q$-index for stars in the saturated regime corresponding to 1, $q=$3 is related to Skumanich index and $q=$5 for the very-low mass stars as propused by Reiners \& Mohanthy (2012). $P_{0}$=1.1 days was used for solar calibration. The Sun is indicated by symbol $\odot$.}
	\label{fig1c}
\end{figure*}

\subsection{Weak version}

A star is considered perfectly spherical when its rotational velocity is zero. In fact, rotating stars with zero velocities, even under the effect of a sighting angle, are rare, and therefore, any star with a velocity greater than zero is a Maclaurin rotation ellipsoid, where the axis of rotation coincides with the shorter axis passing through its centre of mass. A star may remain at a constant volume over time and even exhibit a marked oblateness. In this case, the star is regarded as an incompressible fluid. In general terms, the oblateness depends on the density and the balance of gravitational and centrifugal forces.

From McCoullough's formula, we know that the oblateness can be written as
\begin{equation}
\label{cap52}
\varepsilon=\frac{I-I_{0}}{I_{0}},
\end{equation}
where, $I_{0}$ is the non-rotating spherical moment of inertia and $I$ is the moment of inertia for the rotating star. As mentioned by \cite{gizon}, one mechanism that must be present is rotational oblateness, which is relatively easy to compute when rotation is slow. The centrifugal force distorts the equilibrium structure of a rotating star. The corresponding perturbation of the mode frequencies scales as the ratio of the centrifugal to gravitational forces. Thus, for slow rotators, oblateness is described by a quadrupole distortion of the stellar structure \citep{gizon}. The rotational oblateness $\varepsilon$ depends on the ratio between the centrifugal and gravitational accelerations, as described below.
\begin{equation}
\label{cap53}
\varepsilon=\frac{R^{3}\Omega^{2}}{GM},
\end{equation}
thus, the ``degree'' of oblateness is proportional to be the square of the rotation velocity.  Compiling the equations \ref{cap52} e \ref{cap53}, we find that the moment of inertia, as a function of the rotational velocity of the star, is given by
\begin{equation}
\label{cap58}
I=I_{0}\left(1+\frac{R^{3}\Omega^{2}}{GM}\right),
\end{equation}
where $I$ explicitly depends on $\Omega^{2}$. Then, we have
\begin{equation}
\label{cap510}
\dot{I}=\frac{\mathrm dI}{\mathrm d\Omega}\frac{\mathrm d\Omega}{\mathrm dt}.
\end{equation}
where, from eq. \ref{cap58}, we find that $\frac{\mathrm dI}{d\Omega}$ is
\begin{equation}
\label{cap511}
\frac{\mathrm dI}{\mathrm d\Omega}=2I_{0}\frac{R^{3}\Omega}{GM},
\end{equation}
and therefore,
\begin{equation}
\label{cap512}
\dot{I}=2 I_{0}\frac{R^{3}\Omega}{GM} \dot\Omega.
\end{equation}

Thus, the momentum angular loss rate (see eq. \ref{cap51}) can be rewritten as
\begin{equation}
\label{cap514}
\dot{J}=I_{0}\dot{\Omega}\left(1+\frac{3R^{3}\Omega^{2}}{GM}\right) .
\end{equation}
Let us assume that the momeuntum angular loss rate is a function of only the rotational velocity $\Omega$ and decays as a power law given by
\begin{equation}
\label{cap515}
\dot{J}=K_{m}\Omega^{m},
\end{equation}
where $m$ is a constant and $K_{m}$ is a coefficient that is related to model which describe the angular momemtum loss law. We will consider the coefficient $K_{m}$ constant in time.

In the literature, there are several parametrizations for angular momentum loss through magnetized wind. Among them, the prescriptions of \cite{kawaler1988}, \cite{reiners2012} and \cite{matt} have been widely tested. In each case, there is a functional form for $K_{m}$. For the non-conservative volume case, the term $K_{m}$ will be taken amount. However, in the present case, it is not necessary.	

Combining eqs. \ref {cap514} and \ref{cap515}, we obtain that the first-order derivative of $\Omega$ is given by
\begin{equation}
\label{cap516}
\dot{\Omega}=\frac{K_{m}}{I_{0}}\left(\frac{\Omega^{m}}{1+\frac{3R^{3}\Omega^{2}}{GM}} \right).
\end{equation}

Similarly, by deriving the above equation, we can find the second-order derivative of $\Omega$:
\begin{equation}
\label{cap517}
\ddot{\Omega}=\frac{K_{m}}{I_{0}}\left(\frac{\Omega^{m-1}\dot{{\Omega}}}{1+\frac{3R^{3}\Omega^{2}}{GM}}\right)\left(m-\frac{\frac{6R^{3}\Omega^{2}}{GM}}{1+\frac{3R^{3}\Omega^{2}}{GM}} \right).
\end{equation}

Then, we find the following expression for the magnetic braking index as a function of $\Omega$:
\begin{equation}
\label{cap518}
q(\Omega)=m-\frac{\frac{6R^{3}\Omega^{2}}{GM}}{1+\frac{3R^{3}\Omega^{2}}{GM}}.
\end{equation}

Effectively, the main parameter is the difference between the braking indexes in both cases, considering $I$ as a constant and $\dot{I}\neq0$, i.e., 
\begin{equation}
\label{cap519}
\Delta q=m-q(\Omega)=q(\dot{I}=0)-q(\dot{I}\neq 0).
\end{equation}

Therefore, the contribution of oblateness to the variation of the magnetic braking index can be written as
\begin{equation}
\label{cap519b}
\Delta q=\frac{\frac{6R^{3}\Omega^{2}}{GM}}{1+\frac{3R^{3}\Omega^{2}}{GM}}.
\end{equation}

It is clear from the equation above that only stars with no rotation do not have variation in their braking index. Undoubtedly, the most widespread rotational decay exponent in the literature is that of \cite{sku}. If we do not consider the variation of the moment of inertia, $q(\dot{I}=0)$ is also 3 in equation (\ref{cap519}) and, p. ex., the Skumanich relation will be valid along the main sequence. Any changes of $I$ with velocity $\Omega$ will imply a magnetic braking index $q<3$. In summary, $\Delta q$ is equal to zero when the effect of $\dot{I}$ is neglected. In this case, the oblateness effect is omitted. 

Initially, we can rewrite equation (\ref{cap519b}) in the solar units as
\begin{equation}
\label{cap519b1}
\Delta q=\frac{6\xi \left(\frac{\Omega}{\Omega_{\odot}}\right)^{2}}{1+3\xi \left(\frac{\Omega}{\Omega_{\odot}}\right) ^{2}},
\end{equation}
with
\begin{equation}
\label{cap519b2}
\xi(M,R)=\varepsilon_{\odot}\left( \frac{R}{R_{\odot}}\right) ^{3} \left( \frac{M}{M_{\odot}}\right) ^{-1},
\end{equation}
where the solar oblateness, $\varepsilon_{\odot}$, is $2.14\times 10^{-5}$, assuming $\Omega_{\odot}=2.9\times 10^{-6}$ s$^{-1}$ ($P_{\odot}$=25.4 days), $R_{\odot}=6.96\times 10^{8}$ m, $M_{\odot}=1.99\times 10^{30}$ kg and the gravitational constant $G=6.67\times 10^{-11}$ Nm$^{2}$kg$^{-2}$. For the conservative volume case, we can rewrite the above equation in terms of stellar mean density $\rho$,
\begin{equation}
\label{cap519b21}
\xi(\rho)=\varepsilon_{\odot} \frac{\rho_{\odot}}{\rho},
\end{equation}
where $\rho_{\odot}=1.41$ g/cm$^{3}$ is the solar mean density. 

The equations and procedures listed above are very similar to those adopted by \cite{Yue} for the case of the braking index, which describes the spin-down of pulsars. However, these authors use other parameters of electromagnetic and geometric origin to explain the spin-down, such as the inclination angle, magnetic field and magnetic moment. 

Generally speaking, the stellar mean density decreases as the mass increases (from M to O stars). Eq. (\ref{cap519b1}) reveals that a greater variation of the magnetic braking index occurs in massive stars. This fact is expected since more massive stars are fast rotators, and therefore, the effect of a change of the moment of inertia on these more massive stars is more pronounced. Nevertheless, this effect can be omitted for all the solar-type stars (slow rotators).

The relationship between the $q$-index and $\Omega$ expressed by eq. (\ref{cap519b1}) allows us to state that fast rotators have stronger variations of their magnetic braking ($\Delta q>0$). This conclusion agrees well with the anomalously rapid rotation found in old field stars. Consequently, high values of $\Delta q$ indicate that magnetic braking becomes weaker \citep{van}. After all, what is the meaning of a weakened magnetic braking? For $\Delta q$=0, the magnetic braking is constant over time. For instance, for stars that obey Skumanich law, only the braking exponent $q=3$ would be necessary to describe the rotational behaviour of the star throughout its life in the main sequence. In the opposite situation, if the index differs from zero, we can infer that the magnetic braking was weakened, i.e., for a given value of $q(\dot{I}=0)$, the value of  $q(\dot{I}\neq 0)$ is lower than $q(\dot{I}=0)$, and therefore, the action of the brake is reduced and the rotational velocity would have a value higher than that of the case when $\Delta q=0$.

\begin{figure*}
	\begin{center}
		\includegraphics[width=0.99\textwidth]{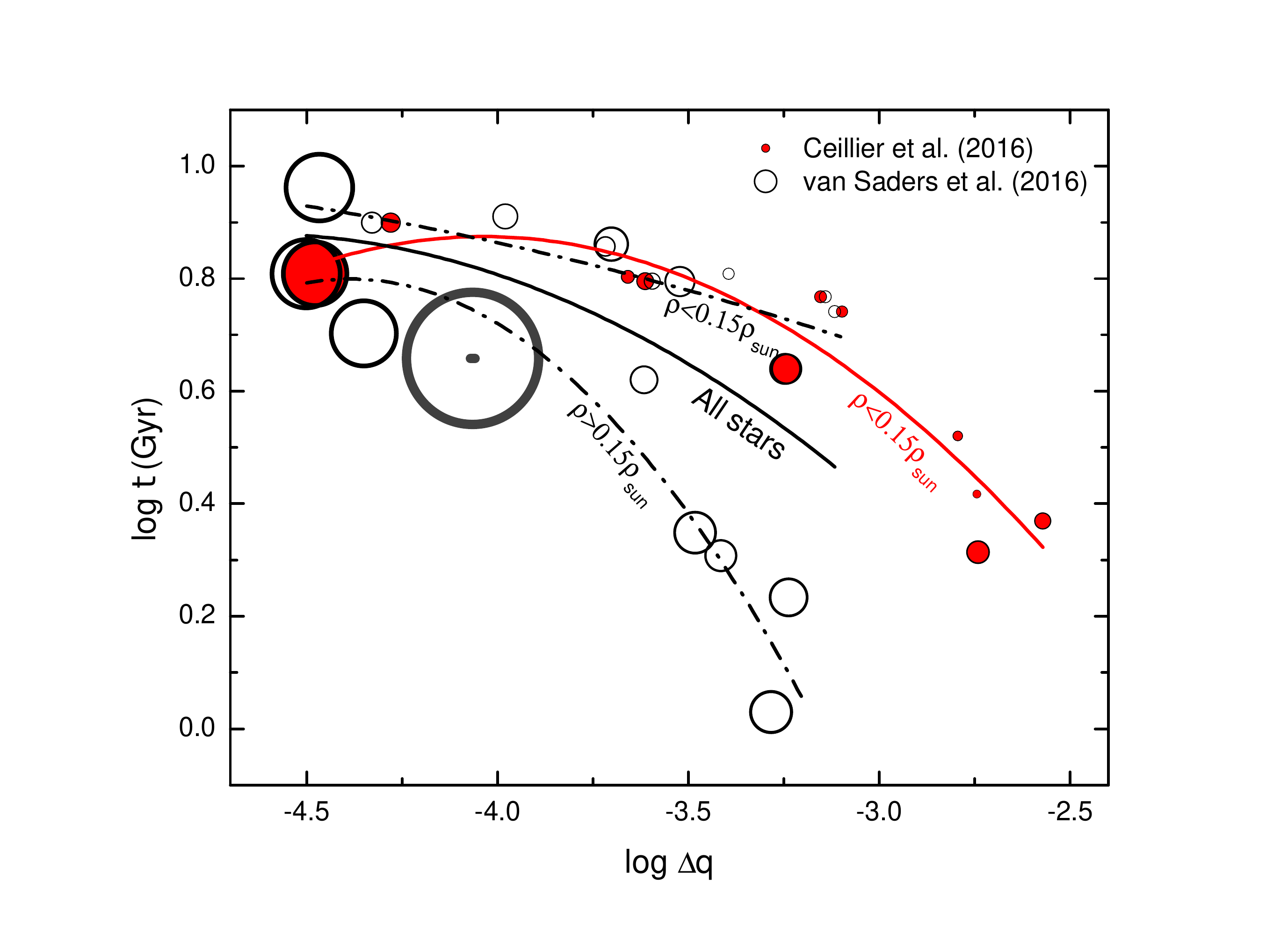}
	\end{center}
	\caption{Logarithmic plot of two samples of stars with well-defined ages, displaying a relationship between the age and the variation of the braking index. Each sample is represented by bubbles with different color and sizes which corresponds to the density of each star within the sample. Non-linear best adjustment for the quadratic model using a similar equations \ref{eq14} to \ref{eq15}. Curves were made following the range of density indicated in figure. The Sun is represented by symbol $\odot$ and it is not used for the adjustment. The diameter of the circles is proportional to the density.}
	\label{fig4}
\end{figure*}

\subsection{Stronger version}

As seen in previous Section, the weak version considers that the changes of $I$ are small and its effect can be neglected when stars have mean density above 10$^{-3}\rho_{\sun}$. In the present case, we will show that a more process effective is possible considering that $\dot{I}$ changes with the radius. It easy to see that the weakness of $\Delta q$ in eq. (\ref{cap511}) is related to term $\frac{\mathrm dI}{\mathrm d\Omega}$ is a function of $\Omega$. 

\begin{equation}
\label{sv1}
\frac{\mathrm dI}{\mathrm d\Omega}=\beta I
\end{equation}
and, therefore
\begin{equation}
\label{sv2}
\frac{\mathrm dI}{\mathrm dt}=\frac{\mathrm dI}{\mathrm d\Omega}\dot{\Omega}=\beta I\dot{\Omega}
\end{equation}
where $\beta$ is a dimensionless coefficient. Replacing in eq. (\ref{cap51}), we have
\begin{equation}
\label{sv3}
\dot{J}=I\dot{\Omega}(1+\beta\Omega)=K_{m}\Omega^{m}.
\end{equation}

Thus, we can derive that 
\begin{equation}
\label{sv4}
\dot{\Omega}=\frac{K_{m}}{I}\frac{\Omega^{m}}{1+\beta\Omega}
\end{equation}
and
\begin{equation}
\label{sv5}
\dot{\Omega}=\frac{K_{m}}{I}\frac{\Omega^{m}}{1+\beta\Omega}\frac{\dot{\Omega}}{\Omega}\left(m-\frac{\beta\Omega}{1+\beta\Omega}\right)
\end{equation}

Acoording to eq. (\ref{4.1}), the magnetic braking $q$ is a function of $\Omega$ by relation
\begin{equation}
\label{sv6}
q(\Omega)=m-\frac{\beta\Omega}{1+\beta\Omega}
\end{equation}
or, then
\begin{equation}
\label{sv7}
\Delta q=m-q(\Omega)=q(\dot{I}=0)-q(\dot{I}\neq 0)=\frac{\beta\Omega}{1+\beta\Omega},
\end{equation}
where $\beta$ can be extracted from eq. (\ref{sv2}). Integrating, we have that
\begin{equation}
\label{sv8}
\beta=\frac{1}{\Omega}\ln\left(\frac{I}{I_{ZAMS}}\right),
\end{equation}
where, finally
\begin{equation}
\label{sv9}
\Delta q(I)=\frac{\ln\left(\frac{I}{I_{ZAMS}}\right)}{1+\ln\left(\frac{I}{I_{ZAMS}}\right)}.
\end{equation}
This version is considered strong because for any value of $I$ the ratio $\frac{\ln\left(\frac{I}{I_{ZAMS}}\right)}{1+\ln\left(\frac{I}{I_{ZAMS}}\right)}$ is much greater than that found in the weak version. The above equation is valid for estimating the corretion $\Delta q$ for post-main sequence, as well as pre-main sequence stars.

\section{Generalized and corrected gyro-ages}
In this section, we will use the conditions imposed by the gyrochronology and, consequently, estimate the ages in two different situations: i) when generalizing the correlation between the age and the period of rotation by using the scaling factor $q$, and ii) when considering the effect of the $\Delta q$--index on the calculation of the gyro ages.

\subsection{A new way for estimating gyro-ages}
Stellar age is the more important stellar parameter for investigating the spin-downs of stars. We have shown in eq. (\ref{4}) from Section 2 that the spin-down rate is proportional to a power law that decays as a function of the magnetic braking index $q$. In general, the solution of the referred equation for the ages of stars, $t$, is given by
\begin{equation}
\label{sa1}
t_{q}=\frac{1}{q-1}\frac{\Omega}{\dot{\Omega}}\left[1-\left(\frac{\Omega}{\Omega_{0}} \right)^{q-1}  \right] ,\quad q>1
\end{equation}
and
\begin{equation}
\label{sa1x}
t_{1}=\frac{\Omega}{\dot{\Omega}}\ln\left(\frac{\Omega}{\Omega_{0}}\right)  ,\quad q=1
\end{equation}
where $\Omega_{0}$ is the angular velocity at time $t=0$ and $\Omega$ is the velocity at time $t=t_{age}$, i.e., now. As quoted by \cite{defreitas2013}, $q>1$ denotes the unsaturated magnetic regime, whereas $q=1$ is the index associated to the saturated regime. 

A later equation would give the star's ``true'' age if only the magnetic torque is responsible for the star spin-down and the $q$-index is a constant. Basically, this generalization is a similar age-dependent factor $g(t)$ in eq. (\ref{int1}), described by a power law of $t^{n}$ where $n=1/(q-1)$.

The solution was only possible because we have assumed that $\lambda_{q}$ and the $q$-index remain constant with time, i.e., the star's moment of inertia does not change over time. In addition, it is easy to show that the stellar age (see eq. \ref{sa1}) increases monotonically with decreasing $q$ (including the special case of $q=1$). As quoted by \cite{defreitas2013}, according to the stellar mass, larger values of $q$ mean that the star spins down more slowly.

A good test is to deduce the gyrochronology ages extracted from \cite{barnes2010a}. These ages can be recovered using the above expression (\ref{sa1}) and eq. (2) from \cite{barnes2010a}'s paper, given by
\begin{equation}
\label{sa2a}
\frac{P}{\dot{P}}=\left(\frac{k_{I}P^{2}}{\tau}+\frac{\tau}{k_{C}}\right),
\end{equation}
where $\tau$ is the convective turnover timescale, and $k_{C}$ and $k_{I}$ are two dimensionless constants whose values can be derived from observations.

To this end, we use the Skumanich index, $q=3$ and $\frac{\Omega}{\dot{\Omega}}=-\frac{P}{\dot{P}}$. Thus, we found that
\begin{equation}
\label{sa2}
t_{q=3}=\left(\frac{\tau}{k_{C}}\right)\left[\frac{1}{2}\left(1-\frac{P^{2}_{0}}{P^{2}}\right)\right]+\left(\frac{k_{I}}{2\tau}\right)\left(P^{2}-P^{2}_{0}\right),
\end{equation}
where this equation is similar to equation (32) from \cite{barnes2010a} 
\begin{equation}
\label{barnes}
t=\left(\frac{\tau}{k_{C}}\right)\ln\left(\frac{P}{P_{0}}\right)+\left(\frac{k_{I}}{2\tau}\right)\left(P^{2}-P^{2}_{0}\right).
\end{equation}

Figure \ref{fig1c} illustrates the behaviours between these two equations, and is clear they are similar from $\sim$ 5 days shown in curves for $q=3$ and Gyro-ages, respectively. In fact, the equations differ only in the first term on the right side. Compared to the second term (where both equations are equal), the first term has a very small contribution, affecting only short periods. In this way, the profiles of the curves $q=3$ and the Gyro-age shown in Fig. \ref{fig1c} are dominated exclusively by the second term. 

On the other hand, using equation (\ref{sa1x}), we can obtain the following expression for stars in the saturated magnetic field regime \citep{chaboyer1995} 
\begin{equation}
\label{sa3}
t_{q=1}=\frac{P}{\dot{P}}\ln\left(\frac{P}{P_{0}}\right),
\end{equation}
where $\frac{P}{\dot{P}}$ is given by eq. (\ref{sa2a}). Therefore, we have
\begin{equation}
\label{sa4}
t_{q= 1}=\left(\frac{k_{I}P^{2}}{\tau}+\frac{\tau}{k_{C}}\right)\ln\left(\frac{P}{P_{0}}\right).
\end{equation}

\begin{longtable}{lccccr}
	\caption{Rotation periods and densities extracted from asteroseismic data, and variation of the braking index and ages obtained from our model.} \label{tab1}\\
	\hline
	Star	&	$P_{rot}$	&	$\rho/\rho_{\odot}$	&	$\log\Delta q$	&	$t_{\Delta q}$	&	Ref.	\\
	&	days	&	days	&	&	Gyr		&		\\
	\hline
	\endhead
	\hline
	\endfoot
	55Cnc	&	39	$\pm$	9	&	1.084	$\pm$	0.04	&	-4.51	$\pm$	0.74	&	7.27	$\pm$	0.98	&	\footnotemark[1]	\\
	CoRoT2	&	4.52	$\pm$	0.02	&	1.362	$\pm$	0.064	&	-2.74	$\pm$	1.35	&	0.84	$\pm$	1.2	&		\\
	CoRoT4	&	8.87	$\pm$	1.12	&	0.79	$\pm$	0.08	&	-3.08	$\pm$	1.03	&	1.65	$\pm$	1.1	&		\\
	CoRoT6	&	6.4	$\pm$	0.5	&	0.929	$\pm$	0.064	&	-2.87	$\pm$	1.19	&	1.19	$\pm$	0.6	&		\\
	CoRoT7	&	23.64	$\pm$	3.62	&	1.671	$\pm$	0.073	&	-4.26	$\pm$	0.84	&	4.41	$\pm$	0.7	&		\\
	CoRoT13	&	13	$\pm$	4	&	0.526	$\pm$	0.072	&	-3.24	$\pm$	0.92	&	2.42	$\pm$	0.2	&		\\
	CoRoT18	&	5.4	$\pm$	0.4	&	1.09	$\pm$	0.16	&	-2.79	$\pm$	0.86	&	1.01	$\pm$	0.56	&		\\
	HATP11	&	30.5	$\pm$	4.5	&	2.415	$\pm$	0.097	&	-4.64	$\pm$	0.01	&	5.68	$\pm$	0.67	&		\\
	HAT21	&	15.9	$\pm$	0.8	&	0.7	$\pm$	0.15	&	-3.54	$\pm$	0.77	&	2.96	$\pm$	0.34	&		\\
	HATS2	&	24.98	$\pm$	0.04	&	1.22	$\pm$	0.06	&	-4.17	$\pm$	1.06	&	4.66	$\pm$	0.56	&		\\
	HD189733	&	11.95	$\pm$	0.01	&	1.98	$\pm$	0.17	&	-3.74	$\pm$	1.18	&	2.23	$\pm$	0.23	&		\\
	HD209458	&	10.65	$\pm$	0.75	&	0.733	$\pm$	0.008	&	-3.21	$\pm$	2.01	&	1.98	$\pm$	0.65	&		\\
	Kepler17	&	12.1	$\pm$	1.56	&	1.121	$\pm$	0.02	&	-3.51	$\pm$	1.82	&	2.25	$\pm$	1.4	&		\\
	Kepler30	&	16	$\pm$	0.4	&	1.42	$\pm$	0.07	&	-3.85	$\pm$	1.35	&	2.98	$\pm$	1.2	&		\\
	Kepler63	&	5.4	$\pm$	0.01	&	1.345	$\pm$	0.083	&	-2.88	$\pm$	1.24	&	1.01	$\pm$	0.99	&		\\
	Qatar2	&	11.4	$\pm$	0.5	&	1.591	$\pm$	0.016	&	-3.61	$\pm$	2.10	&	2.12	$\pm$	0.78	&		\\
	WASP4	&	22.2	$\pm$	3.3	&	1.23	$\pm$	0.022	&	-4.07	$\pm$	1.61	&	4.14	$\pm$	0.45	&		\\
	WASP-5	&	16.2	$\pm$	0.4	&	0.801	$\pm$	0.08	&	-3.61	$\pm$	1.10	&	3.02	$\pm$	0.32	&		\\
	WASP-10	&	11.91	$\pm$	0.05	&	2.359	$\pm$	0.05	&	-3.81	$\pm$	1.81	&	2.22	$\pm$	0.12	&		\\
	WASP19	&	11.76	$\pm$	0.09	&	0.885	$\pm$	0.006	&	-3.38	$\pm$	2.24	&	2.19	$\pm$	0.02	&		\\
	WASP41	&	18.41	$\pm$	0.05	&	1.27	$\pm$	0.14	&	-3.92	$\pm$	0.98	&	3.43	$\pm$	0.34	&		\\
	WASP46	&	16	$\pm$	1	&	1.24	$\pm$	0.1	&	-3.79	$\pm$	1.18	&	2.98	$\pm$	0.59	&		\\
	WASP50	&	16.3	$\pm$	0.5	&	1.376	$\pm$	0.032	&	-3.85	$\pm$	1.70	&	3.04	$\pm$	0.63	&		\\
	WASP69	&	23.07	$\pm$	0.16	&	1.54	$\pm$	0.13	&	-4.20	$\pm$	0.77	&	4.30	$\pm$	0.25	&		\\
	WASP77	&	15.4	$\pm$	0.4	&	1.157	$\pm$	0.018	&	-3.73	$\pm$	1.90	&	2.87	$\pm$	0.29	&		\\
	WASP84	&	14.36	$\pm$	0.35	&	2.015	$\pm$	0.07	&	-3.91	$\pm$	1.53	&	2.68	$\pm$	0.54	&		\\
	WASP85	&	14.64	$\pm$	1.47	&	1.28	$\pm$	0.01	&	-3.73	$\pm$	2.16	&	2.73	$\pm$	0.41	&		\\
	WASP89	&	20.2	$\pm$	0.4	&	1.357	$\pm$	0.07	&	-4.03	$\pm$	1.22	&	3.76	$\pm$	0.6	&		\\
	\hline
	KIC3427720	&	13.9	$\pm$	2.1	&	0.192	$\pm$	0.052	&	-3.48	$\pm$	0.57	&	2.59	$\pm$	0.04	& \footnotemark[2]	\\
	KIC3656476	&	31.7	$\pm$	3.5	&	0.116	$\pm$	0.035	&	-3.98	$\pm$	0.52	&	5.91	$\pm$	0.33	&		\\
	KIC5184732	&	19.8	$\pm$	2.4	&	0.129	$\pm$	0.040	&	-3.62	$\pm$	0.51	&	3.69	$\pm$	0.09	&		\\
	KIC6116048	&	17.3	$\pm$	2	&	0.136	$\pm$	0.046	&	-3.52	$\pm$	0.47	&	3.22	$\pm$	0.07	&		\\
	KIC6196457	&	16.4	$\pm$	1.2	&	0.059	$\pm$	0.023	&	-3.12	$\pm$	0.42	&	3.06	$\pm$	0.03	&		\\
	KIC6521045	&	25.3	$\pm$	2.8	&	0.075	$\pm$	0.023	&	-3.59	$\pm$	0.51	&	4.72	$\pm$	0.11	&		\\
	KIC7680114	&	26.3	$\pm$	1.9	&	0.092	$\pm$	0.034	&	-3.72	$\pm$	0.43	&	4.90	$\pm$	0.18	&		\\
	KIC7871531	&	33.7	$\pm$	2.6	&	0.315	$\pm$	0.097	&	-4.47	$\pm$	0.51	&	6.28	$\pm$	1.15	&		\\
	KIC8006161	&	29.8	$\pm$	3.1	&	0.309	$\pm$	0.087	&	-4.35	$\pm$	0.55	&	5.55	$\pm$	0.71	&		\\
	KIC8349582	&	51	$\pm$	1.5	&	0.100	$\pm$	0.037	&	-4.33	$\pm$	0.43	&	9.51	$\pm$	1.47	&		\\
	KIC9098294	&	19.8	$\pm$	1.3	&	0.157	$\pm$	0.055	&	-3.70	$\pm$	0.45	&	3.69	$\pm$	0.13	&		\\
	KIC9139151	&	11	$\pm$	2.2	&	0.174	$\pm$	0.042	&	-3.24	$\pm$	0.61	&	2.05	$\pm$	0.02	&		\\
	KIC9955598	&	34.7	$\pm$	6.3	&	0.322	$\pm$	0.083	&	-4.50	$\pm$	0.59	&	6.47	$\pm$	1.07	&		\\
	KIC10454113	&	14.6	$\pm$	1.1	&	0.149	$\pm$	0.037	&	-3.41	$\pm$	0.61	&	2.72	$\pm$	0.03	&		\\
	KIC10586004	&	29.8	$\pm$	1	&	0.063	$\pm$	0.027	&	-3.66	$\pm$	0.37	&	5.55	$\pm$	0.21	&		\\
	KIC10644253	&	10.91	$\pm$	0.87	&	0.197	$\pm$	0.058	&	-3.28	$\pm$	0.53	&	2.03	$\pm$	0.02	&		\\
	KIC10963065	&	12.4	$\pm$	1.2	&	0.139	$\pm$	0.040	&	-3.24	$\pm$	0.54	&	2.31	$\pm$	0.02	&		\\
	KIC11244118	&	23.2	$\pm$	3.9	&	0.056	$\pm$	0.020	&	-3.39	$\pm$	0.45	&	4.32	$\pm$	0.07	&		\\
	KIC11401755	&	17.2	$\pm$	1.4	&	0.057	$\pm$	0.014	&	-3.14	$\pm$	0.60	&	3.21	$\pm$	0.02	&		\\
	\hline
	KIC3632418	&	12.5	$\pm$	1.18	&	0.049	$\pm$	0.008	&	-2.79	$\pm$	0.77	&	3.24	$\pm$	0.45	&	\footnotemark[3]	\\
	KIC5866724	&	7.89	$\pm$	0.68	&	0.108	$\pm$	0.029	&	-2.74	$\pm$	0.58	&	1.37	$\pm$	0.46	&		\\
	KIC6196457	&	16.42	$\pm$	1.22	&	0.057	$\pm$	0.011	&	-3.10	$\pm$	0.70	&	3.94	$\pm$	0.32	&		\\
	KIC6521045	&	25.34	$\pm$	2.78	&	0.078	$\pm$	0.009	&	-3.61	$\pm$	0.93	&	5.18	$\pm$	0.49	&		\\
	KIC8349582	&	51.02	$\pm$	1.45	&	0.089	$\pm$	0.040	&	-4.28	$\pm$	0.35	&	9.75	$\pm$	0.4	&		\\
	KIC9592705	&	13.41	$\pm$	1.11	&	0.038	$\pm$	0.007	&	-2.74	$\pm$	0.72	&	3.95	$\pm$	0.43	&		\\
	KIC9955598	&	34.75	$\pm$	6.31	&	0.304	$\pm$	0.058	&	-4.48	$\pm$	0.72	&	3.60	$\pm$	0.35	&		\\
	KIC10586004	&	29.79	$\pm$	1.02	&	0.063	$\pm$	0.015	&	-3.66	$\pm$	0.63	&	6.80	$\pm$	0.9	&		\\
	KIC10963065	&	12.38	$\pm$	1.22	&	0.141	$\pm$	0.022	&	-3.25	$\pm$	0.81	&	1.89	$\pm$	0.39	&		\\
	KIC11401755	&	17.04	$\pm$	0.98	&	0.060	$\pm$	0.025	&	-3.15	$\pm$	0.37	&	3.98	$\pm$	1.29	&		\\
	KIC11807274	&	7.71	$\pm$	0.66	&	0.077	$\pm$	0.018	&	-2.57	$\pm$	0.63	&	1.59	$\pm$	0.22	&		\\
\end{longtable}
\footnotetext[1]{\cite{max}.}
\footnotetext[2]{\cite{van}.}
\footnotetext[3]{\cite{cei}.}

\subsection{Effect of $\Delta q$ on the gyro--ages}
In the previous section, we saw that a $q$-index is a constant. On the other hand, if we assume $q$ as a function of $\Omega$ (see eqs. \ref{cap518} and \ref{sv6}) the gyro ages of \cite{barnes2010a} should be revisited. Thus, when $q$ varies over $\Omega$, equation (\ref{4}) can be generalized as
\begin{equation}
\begin{array}{cc}
\dot{\Omega}=-\lambda _{q}\Omega^{q(\Omega)}; & (\lambda _{q}\geqslant 0;q\geqslant 1),
\end{array}
\label{4a}
\end{equation}
which, when solved, would give the star's ``true age''. Again, as a consequence of its dependence on $\Omega$, the braking index $q$ is not constant in time. In this context, the present section will show the effects of $\Delta q$ on the calculations of the gyro ages. For gyrochronological model $q(\dot{I}=0)=3$ and, therefore, $q(\Omega)=3-\Delta q$. On the other hand, $-\lambda _{q}$ can be written as
\begin{equation}
-\lambda_{q}=\left( \frac{\dot{\Omega}}{\Omega}\right)\Omega^{\Delta q-2},
\label{sa6ax}
\end{equation}

For this purpose, let us write the new gyro--age corrected by factor $\Delta q(\Omega)$, $t_{\Delta q}$, from eqs. (\ref{4a}) and (\ref{sa6ax}) in the form 
\begin{equation}
t_{\Delta q}=-\left( \frac{\Omega}{\dot{\Omega}}\right)\Omega^{2-\Delta q}\int^{\Omega}_{\Omega_{0}} \frac{d\Omega}{\Omega^{3-\Delta q}}.
\label{sa6a}
\end{equation}
or, then
\begin{equation}
t_{\Delta q}=-\left(\frac{4\pi^{2}k_{I}}{\tau\Omega^{2}}+\frac{\tau}{k_{C}}\right)\Omega^{2-\Delta q}\int^{\Omega}_{\Omega_{0}} \frac{d\Omega}{\Omega^{3-\Delta q}},
\label{sa6a}
\end{equation}
where $\Delta q$ is given by two forms: i) $\Delta q_{W}$ defined by Weak version (see eq.~\ref{cap519b1}) and ii) $\Delta q_{S}$ defined by Strong version (see eq.~\ref{sv9}). Equation~\ref{sa6a} is calculated numerically when $\Delta q=\Delta q_{W}$. On the other hand, it is solved analycally when $\Delta q=\Delta q_{S}$, hence
\begin{equation}
\label{sa2S}
t_{\Delta q_{S}}=\frac{1}{1-\Delta q_{S}/2}\left \{\left(\frac{\tau}{k_{C}}\right)\left[\frac{1}{2}\left(1-\frac{P^{2-\Delta q_{S}}_{0}}{P^{2-\Delta q_{S}}}\right)\right]+\left(\frac{k_{I}}{2\tau}\right)\left[P^{2}-P^{2}_{0}\left(\frac{P}{P_{0}}\right)^{\Delta q_{S}}\right]\right \},
\end{equation}
here, $t_{\Delta q_{S}}$ is valid for $0\leq\Delta q_{S}<2$. Is is clear that for $\Delta q_{S}=0$ the gyro-ages are recovered.

We solve this differential equation (right side from eq. \ref{sa6a}) numerically, fixing the velocity $\Omega$, and find a solution depending on two parameters: one unknowns defined above as the initial angular velocity $\Omega_{0}$, and one known as $\xi(M,R)$. 

We estimated the uncertainties of all the parameters (e.g., $\Delta q$ and $t_{\Delta q}$) using a general equation given by expression 3.47 (p.75) from \cite{taylor}. Initially, we have the errors of the radius, mass and period obtained from the selected samples. The other parameters such as the density, $\Delta q$ and age, we derive through the propagation of the uncertainties following the methodology employed by \cite{taylor}. In the next section, we tested our model by deriving the ages and comparing these ages with those by different methods, such as asteroseismology, isochronology and gyrochronology.

Our technique of $\Delta q$-ages uses  the rotation period of a star to calculate its age and uses the density of a star as the parameter that segregates ages in different regimes, as can be seen in Fig.~\ref{fig4}. Consequently, if the sample has a density that are not compatible with the density of the Sun, the solar value cannot be used for the calibration of the age and, therefore, stars of known ages should be used for this purpose.

As shown in figure~\ref{fig1c}, the effect of the variation of the braking index is more accentuated in more massive stars with a period of rotation of less than 20 days. It is evident that the theoretical predictions of the ages of stars considering. For example, the Skumanich law without the correction of the variation of the braking index gives an overestimated value for the stellar age using the same conditions for $\Omega_{0}$, $\rho$ and $q=q(\dot{I}=0)$. This difference is caused by the greater contribution of the braking index as the stellar mass grows. Therefore, a difference of this order warns us that the estimated ages for stars that do not consider the effects of the changes of moments of inertia can generate considerable errors. The braking indices found in figure~\ref{fig1c} emphasize some values of $q$ found in the literature, from stars in the saturated regime of the magnetic field ($q=1$) to very low mass stars ($q=5$), as investigated by \cite{reiners2012} and \cite{defreitasetal15}. In the same figure, we assumed the adjustment recommended by \cite{barnes2010a} to calibrate the present model with the Sun. The model also reveals that cool stars do not suffer greater effects of the variations of the magnetic braking index over time.

\begin{figure*}
	\begin{center}
		\includegraphics[width=0.99\textwidth]{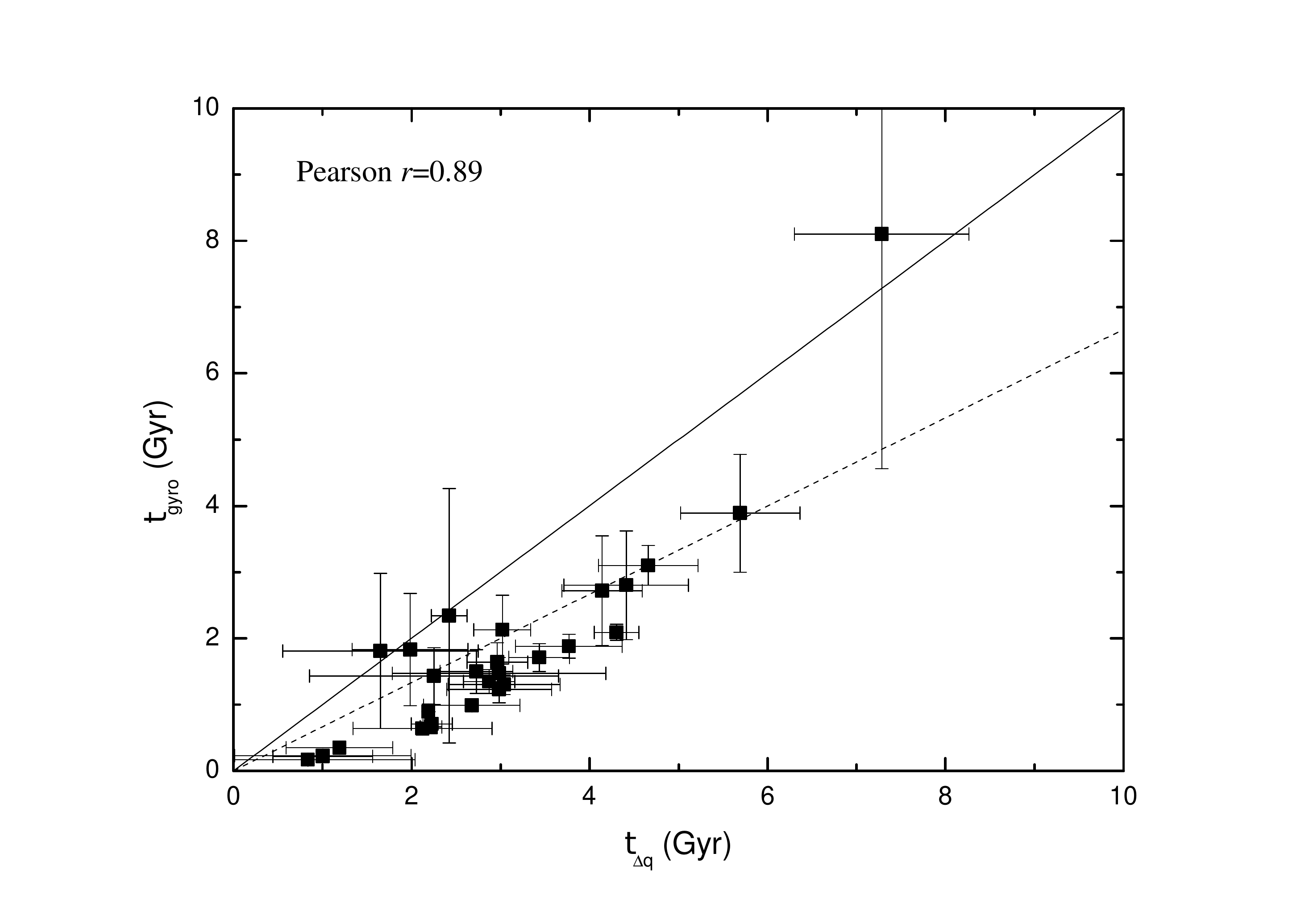}
	\end{center}
	\caption{Comparison of asteroseismic ages ($t_{gyro}$) to $\Delta q$--ages ($t_{\Delta q}$) for Maxted, Serenelli \& Southworth' sample. The straight line is the relation $t_{\Delta q}=t_{gyro}$.}
	\label{figAges3}
\end{figure*}

\section{Comparing derived ages from our model}
To compare the ages calculated by our model, we chose different samples that use different methods to estimate the age of a star. In addition to the two samples already mentioned in the previous section that were obtained through asteroseismology, we included another sample whose ages were measured by isochrones and gyrochronology \citep{max}. We will briefly describe the fundamental features of each sample and their measured ages. Our stellar targets come from different instruments, including the Kepler and Corot spacecrafts. Here, we focused on stars with temperature values below 6200K:

\textit{\cite{van}} -- this sample comprises 21 stars, where 19 are from the Kepler database from quarters 5--17 with a short cadence of $\sim$ 1 min and 2 stars that correspond to the 16 Cygni binary system. However, in the present analysis, we have removed this system. Table 1 from \cite{van} presents the spectroscopic quantities and asteroseismological data, such as the temperature, metallicity, mass and effective gravity. According to this paper, the rotational periods were calculated using the autocorrelation function (ACF) and wavelet techniques, where the peaks of both techniques are compared and confirmed via a visual inspection of the time series. As mentioned in the previous section, the asteroseismological ages ($t_{astero}$) were measured using the BASTA (Bayesian stellar algorithm) and AMP (asteroseismic modeling portal) methods, which, in general, use frequency spacing and spectroscopic constraints to identify the optimal stellar properties. The uncertainties were determined using the distribution of the generated models as described by \cite{van}. The stellar radii and their uncertainties were extracted from \cite{huber}'s catalogue.

\textit{\cite{cei}} -- This sample comprises 11 known planet-hosting stars from the \textit{Kepler} mission (KOIs). For these stars, the ACF and wavelet methods are able to extract a surface rotation period. For the extraction of the rotation period, they used the Kepler long-cadence data for all quarters (from 0 to 17) with a cadence at $\sim$ 30 min. The asteroseismological ages were also calculated using the BASTA and AMP codes. The uncertainties were determined by using the same procedure proposed by \cite{van}. Table 1 from \cite{cei} presents all the stellar parameters and uncertainties used in the present analysis.

\textit{\cite{max}} -- This sample comprises 28 transiting exoplanet host stars with measured rotational periods from different instruments, including \textit{Kepler}, CoRoT and WASP. The authors use a new approach based on the Bayesian Markov chain Monte Carlo (MCMC) method to determine the joint posterior distribution for the mass and age of each star in the sample and their uncertainties. The authors offer two measured ages by using the isochrone and gyrochronology methods. In addition, they used equation (32) from \cite{barnes2010a} to calculate the gyrochronological ages ($t_{gyro}$) from the rotation period and the mass of the star for every point in the Markov chain for each star. The isochronal ages ($t_{iso}$) were derived by comparing the properties of a star to a grid of stellar models, as mentioned by \cite{max}. In addition, the densities were estimated using asteroseismology \citep{huber2013,huber}.

Our procedure uses an algorithm to perform a search for a global $\chi ^{2}$ minimum, which is calibrated to the solar age value of 4.55 Gyr or a star of known age, depending on the density of the star. In our model, there are two free parameters: $q$ and $\lambda_{q}$. The best values of $q$ and $\lambda_{q}$ are shown in Table \ref{tab2}. This Table shows that there is a clear anti-correlation between $\lambda_{q}$ and the mean density $\langle\rho\rangle$.

The figures \ref{figAges1} to \ref{figAges4} show a comparison between the measured ages of different samples collected from the above cited sample and the ages extracted from our model (see Table \ref{tab1} for further details). We find that these ages are compatible with the measured ages of other methods. For \cite{cei}'s sample, the density range is considerably below the solar density. Thus, we choose KIC10963065, which has a near-solar age, as a reference for calculating ages in our method. As a result, the Pearson's coefficients ($r$) were calculated: 0.73 for \cite{van}'s sample, 0.80 for \cite{cei} and 0.89 for the gyro ages of \cite{max}. In contrast, for the isochronal ages of \cite{max}, the correlation was estimated to be very weak. In particular, figure \ref{figAges3} also shows that there is strong evidence that the gyrochronological ages of the transiting exoplanet host stars are significantly less than their $\Delta q$-ages. This difference may be related to tidal interactions between the star and the planet. However, this discrepancy is not as great as that found by \cite{max} when comparing the isochronal and gyrochronological ages.

We also found that the statistical weight of $\Delta q$ on the age determination is greater than the statistical one due to the density, emphasizing that the variation of the braking index is a fundamental factor in our model. We also found that, in all samples, the best value for $q(\dot{I}=0)$ is close to 2, which differs from the Skumanich exponent. We have already found $\lambda_{q}$ to be anti-correlated with mass. 

\begin{longtable}{lccr}
	\caption{Results of statistical test for the parameters 	$q(\dot{I}=0)$,	$\lambda_{q}$ and $\langle\rho\rangle$ for collected samples.} \label{tab2}\\
	\hline
	Sample	&	$q(\dot{I}=0)$	&	$\lambda_{q}$	& $\langle\rho\rangle$\\
	\hline
	\endhead
	\hline
	\endfoot
	\cite{max}	&	2.00	$\pm$	0.01	& 367.04 $\pm$ 6.87	&  1.30\\
	\cite{van}	&	2.01	$\pm$	0.02	& 751.88$\pm$ 7.13	&  0.150\\
	\cite{cei}	&	1.98	$\pm$	0.01	& 107526.9$\pm$ 746.98	&  0.097\\
	
\end{longtable}

\section{Concluding remarks}

Undoubtedly, the strongest conclusion in this study concerns the role of the variation of the magnetic braking ($\Delta q$) as a ``new parameter'' to explain the mechanism that controls stellar clocks. We find that $\Delta q$ is a measurement of the strength of the magnetic braking caused by rotation and density. In this sense, $\Delta q$ is a parameter sensitive to the oblateness effects that determine how strong magnetized winds are. In addition, this parameter provides us with a powerful diagnostic tool concerning the "rotational lives" of the stars on the main sequence. In particular, the weakened magnetic braking as the origin of the anomalously rapid rotation in old field stars, as suggested by \cite{van}, is, in the context of our model, understood as an increase in the broadness of $\Delta q$ and, therefore, should be associated with a change in the moment of inertia of the stars. This is a strong point of our model, in that the model can improve the diagnosis of the transition to weaker magnetized winds.

Given the strong agreement of our model with the previously estimated ages of other samples, we suggest that the variation of the magnetic braking index ($\Delta q$) can be an interesting and alternative way to estimate stellar ages. Finally, it is worth mentioning that the $\Delta q$-index can be estimated for other scenarios, including, for instance, globular and open clusters as well as giant stars, where the effects of the variations of stellar radii should be considered. This issue will be addressed in a forthcoming communication.

\acknowledgments
DBdeF acknowledges financial support 
from the Brazilian agency CNPq-PQ2 (grant No. 306007/2015-0). Research activities of STELLAR TEAM of Federal University of Cear\'a are supported by continuous grants from the Brazilian agency CNPq.

\end{document}